\documentstyle[12pt]{article}
\begin{document}
\title{On the flavour dependence of polarized sea in the nucleon
\thanks{Work supported in part by the KBN-Grant PB 2-P03B-065-10.}}
\author{Jan Bartelski\\
Institute of Theoretical Physics, Warsaw University,\\
Ho\.{z}a 69, 00-681 Warsaw, Poland. \\ \\
\and
Stanis\l aw  Tatur \\
Nicolaus Copernicus Astronomical Center,\\ Polish Academy of Sciences,\\
Bartycka 18, 00-716 Warsaw, Poland.\\}
\date{}
\maketitle
\begin{flushright} preprint CAMK No. 328 \end{flushright}
\vspace{2cm}
\begin{abstract}
\noindent We performed a phenomenological fit in order to get quark parton 
polarized distributions in the nucleon. All data on inclusive 
and semi-inclusive spin asymmetries measured on nucleon targets were used.
We present the results for the flavour dependence of polarized sea inside 
a nucleon. An excellent agreement between inclusive and semi-inclusive
data on polarized structure functions was found in our model.
\end{abstract}

\newpage

Recently the SMC collaboration has presented \cite{SMC} the experimental 
results of semi-inclusive spin asymmetries for positively and negatively 
charged hadrons.
A lot of earlier results for the polarized inclusive deep inelastic 
scattering on protons, neutrons ($^{3}He$) and deuterons
\cite{d,emc} are also available. 
There were also theoretical attempts to get the polarized parton 
distributions using these measured asymmetries \cite{brod}.
The SMC group has tried to determine (using inclusive and
semi-inclusive hadron asymmetries) spin distributions 
for valence (up and down) quarks and for non-strange sea quarks. 
In this paper we want to extend our previous determination of polarized 
parton distributions \cite{bt,bt1} obtained from the
inclusive spin asymmetries by taking into account new semi-inclusive
data. With the new hadronic semi-inclusive asymmetries we will
try to get more information about flavour structure of polarized sea.
Our idea, as before, is to use the unpolarized parton distributions 
(see for example \cite{mrs,mrs2}) and to split $ q(x)$
($q(x)=q^{+}(x)+q^{-}(x)$) in order to get the polarized ones, i.e.  
$q^{+}(x)$ and $q^{-}(x)$. With such a procedure one has simultaneous 
description of spin and spin averaged structure functions.

The latest version of the MRS (Martin, Roberts and Stirling 
\cite{mrs2}) fit to unpolarized parton distributions yields
for the valence quarks (at $ Q^{2}=4\, {\rm GeV^{2}}$):
\begin{eqnarray}
u_{v}(x)&=&2.704 x^{-0.407}(1-x)^{3.96}(1-0.76\sqrt{x}+4.20x),
\nonumber \\
d_{v}(x)&=&0.251 x^{-0.665}(1-x)^{4.41}(1+8.63\sqrt{x}+0.32x) ,
\end{eqnarray}

\noindent whereas for the antiquarks from the sea one has:
\begin{eqnarray}
2\bar{u} (x)&=&0.392M(x)-\delta (x), \nonumber \\
2\bar{d} (x)&=&0.392M(x)+\delta (x), \nonumber \\
2\bar{s} (x)&=&0.196M(x), \\
2\bar{c} (x)&=&0.020M(x). \nonumber
\end{eqnarray}

\noindent The singlet contribution (which gives the total polarization 
of the nucleon sea) is:
\begin{equation}
M(x)=1.74 x^{-1.067}(1-x)^{10.1}(1-3.45\sqrt{x}+10.3x).
\end{equation}

\noindent For the isovector part of quark sea and for gluon 
distribution we have, respectively:
\begin{equation}
\delta (x)=0.043x^{-0.7}(1-x)^{10.1}(1+64.9x),
\end{equation}

\begin{equation}
G(x)=1.51x^{-1.301}(1-x)^{6.06}(1-4.14\sqrt{x}+10.1x).
\end{equation}

We assume, in an analogy to the unpolarized case, that the polarized
quark distributions are of the form:
$x^{\alpha}(1-x)^{\beta}P_{2}( \sqrt{x})$, where
$P_{2}(\sqrt{x})$ is a second order polynomial in $\sqrt{x}$ and
the asymptotic behaviour for $x$$\rightarrow$$0$ and
$x$$\rightarrow$$1$ (i.e. the values of $\alpha$ and $\beta$)
are the same (except for $\Delta M$) as in
unpolarized case.
We split the numerical constants
(coefficients of $P_{2}$ polynomial)  in
eqs.(1, 3 and 4)  in two parts in
such a manner that we get positive defined distributions.
Our expressions for $\Delta q(x)$  ($\Delta q(x) = q^{+}(x)-q^{-}(x)$)
for valence quarks are:
\begin{eqnarray}
\Delta u_{v}(x)&=&x^{-0.407}(1-x)^{3.96}(a_{1}+a_{2}\sqrt
{x}+a_{3}x), \nonumber \\
\Delta
d_{v}(x)&=&x^{-0.665}(1-x)^{4.41}(b_{1}+b_{2}\sqrt{x}+b_{3}x).
\end{eqnarray}

\noindent 

For a moment we will not take into account polarized gluons, 
i.e. we put $\Delta G =0$.

In our earlier fits we had assumed that the total sea
polarization ($\Delta M$) has no
term behaving like $x^{-1.067}$ at small $x$ (we assumed that 
all distributions are integrable) and only one (as in unpolarized case) 
is determinig all
sea contributions for different flavours: up, down, strange and
charm. That assumption was used to reduce the number of unknown
parameters but it was very restrictive. The strenght of strange quark sea
polarization $\Delta s$ was practically determined by the value of $a_{8}$.
With  high value of $\Delta s$ (or equivalently $\Delta M$)
$\Delta\bar{u}$ and $\Delta\bar{d}$ were also forced to be big
(especially $\Delta\bar{d}$) because $\Delta \delta$ (see eq.(2)) was negative.
That was rather unnatural. With the additional new data on
semi-inclusive hadron asymmetries we can relax this assumption
and see what are the polarized sea contributions for different
flavours. Now we propose that for different flavours polarized
sea distributions are different (except for the charm flavour but 
this is completely unimportant numerically), namely:
\begin{equation}
\Delta M_{i}(x)=x^{-0.567}(1-x)^{10.1}(c+c_{i}\sqrt{x}).
\end{equation}
In particular we have for sea contribution of different flavours:
\begin{eqnarray}
\Delta{u_M}=\Delta\bar{u} (x)&=&0.196\Delta M_{u}(x)-0.5\Delta \delta (x), 
\nonumber \\
\Delta{d_M}=\Delta\bar{d} (x)&=&0.196\Delta M_{d}(x)+0.5\Delta \delta (x), 
\nonumber \\
\Delta{s_M}=\Delta\bar{s} (x)&=&0.098\Delta M_{s}(x), \\
\Delta{c_M}=\Delta\bar{c} (x)&=&0.010\Delta M_{s}(x), \nonumber
\end{eqnarray}

\noindent where $\Delta{q_M}(x)=\Delta{q}(x)-\Delta{q_v}(x)-
\Delta{\bar{q}}(x)$ and: 
\begin{equation}
\Delta \delta (x)=x^{-0.7}(1-x)^{10.1}d(1+64.9x).
\end{equation}
We assume that the sea
contributions for quarks and antiquarks are equal (this assumption 
will be reconsidered later).
In order to get the unknown parameters in the expressions for polarized
quark distributions at $ Q^{2}=4\, {\rm GeV^{2}}$ (see eqs.(6-9)) we make 
a fit to the experimental data on inclusive spin asymmetries for
proton, neutron and deuteron targets and semi-inclusive hadron
asymmetries on proton and deuteron.
The theoretical expressions for inclusive spin asymmetries are given by:

\begin{equation}
A_{1}(x,Q^2)=\frac{\sum_{q}e_{q}^2\Delta
q(x,Q^2)}{\sum_{q}e_{q}^2 q(x,Q^2)}(1+R)
\end{equation}
where $e_{q}$ is the charge of the q-flavoured quark.
The ratio $R=\sigma_{L}/\sigma_{T}$ (which vanishes in the
Bjorken limit) is taken
from \cite{whit}. The expression for deuteron asymmetry is considered 
to be the
sum of the corresponding expressions for proton and neutron
corrected for the D-state portion in the deuteron ($p_{D}$),
equal to $(5\pm 1)\%$ \cite{d}. 
The expressions for semi-inclusive spin asymmetries for the production of
positive and negative hadrons are given by:
\begin{equation}
A^{+(-)}_{1}(x,Q^2)=\frac{\sum_{q,h}e_{q}^2\Delta
q(x,Q^2) D^{h}_{q}(Q^2)}{\sum_{q,h}e_{q}^2 q(x,Q^2) D^{h}_{q}(Q^2)}(1+R)
\end{equation}
where $D^{h}_{q}(Q^2)=\int_{0.2}^{1} dz D^{h}_{q}(z,Q^2)$.
Following the usual convention all quark distributions refer to
the proton. To reduce the number of independent fragmentation
functions we use the same relations as in ref.\cite{SMC} and
\cite{frag}. From eq.(11) one can see that the presence of different
weights ($D^{h^{\pm}}_{q}$) enables to extract different combinations 
of $\Delta q$ than in the inclusive spin asymmetries.
First of all we investigate what happens when we add
measured semi-inclusive hadron asymmetries to the sample of
inclusive data on spin structure of the nucleon. Taking the
parameters of the fit from our previous paper \cite{bt1} we can
calculate that the contribution of semi-inclusive data from SMC
group gives $\chi^2_{SMC}=62.5$ (for 48 experimental points) and
from EMC \cite{emc} $\chi^2_{EMC}=20.2$ (for 10 points) and together with 
$\chi^2_{incl}=96.5$ we get $\chi^2_{total}\cong 179$. 
We can also compare our predictions for $A^{+(-)}_{1p}$ and
$A^{+(-)}_{1d}$ with the measured SMC and EMC
results. When we make fit to all the data (inclusive and
semi-inclusive) the total value is $\chi^2_{all}=178.5$ so we do not
gain much in $\chi^2$. The fitted parameters are very close to the
parameters obtained with inclusive data only (the new calculated
values for $A^{+(-)}_{1p}$ and $A^{+(-)}_{1d}$ also change very
little.) That means that the new semi-inclusive data are
consistent with inclusive data.
{\em We assume} that the spin asymmetries do not depend on 
$Q^{2}$ (it is only our first order approximation) what is
suggested by the experimental data \cite{d}.
We hope that 
numerically our results  at $ Q^{2}=4\, {\rm GeV^{2}}$ will not change
much if the evolution of $F_1$ and $g_1$ functions will be taken into 
account.
The semi-inclusive hadron asymmetries have
relatively big errors (there are also uncertainties connected
with the quark fragmentation functions) and also it is not clear how
reliable is the fit to R function for small $x$. Taking all that into
account we will take crude approximation neglecting the $Q^{2}$
dependence of the spin asymmetries.

The obtained quark distributions $\Delta u_{v}(x)$,
$\Delta d_{v}(x)$, $\Delta M_{u}(x)$, $\Delta M_{d}(x)$, $\Delta
M_{s}(x)$ and $\Delta\delta(x)$ can be used
to calculate first moments. For a given $Q^{2}$ we can write the
relations:
\begin{eqnarray}
\Gamma^{p}_{1}& = &\frac{4}{18}\Delta u+\frac{1}{18}\Delta d+
\frac{1}{18}\Delta s+\frac{4}{18}\Delta c, \nonumber \\
\Gamma^{n}_{1}& = &\frac{1}{18}\Delta u+\frac{4}{18}\Delta d+
\frac{1}{18}\Delta s+\frac{4}{18}\Delta c ,
\end{eqnarray}

\noindent where $\Delta q=\int^{1}_{0}\Delta q(x)\, dx$ and
$\Gamma_{1}=\int^{1}_{0}g_{1}(x)\, dx$.

We define other combinations of integrated quark polarizations:
\begin{eqnarray}
a_{3}& = &\Delta u-\Delta d , \nonumber \\
a_{8}& = &\Delta u+\Delta d-2\Delta s , \\
\Delta\Sigma& =&\Delta u+\Delta d+\Delta s , \nonumber
\end{eqnarray}

Such results for the integrated quantities (calculated at $4\,{\rm GeV^{2}}$)
after taking into account known QCD corrections (see e.g. Ref.\cite{lar})
could be compared with axial-vector coupling constants
$g_{A}$ and $g_{8}$ known from neutron $\beta$-decay and
hyperon $\beta$-decays (in the last case one needs $SU(3)$
symmetry).
In our paper $Q^2$ is constant and takes the value $4\,{\rm GeV^{2}}$.
Using experimental numbers \cite{cr} we expect that $a_{3}(4\,{\rm GeV^{2}})=
1.11$ and $a_{8}(4\,{\rm GeV^{2}})=0.51\pm 0.03$.
As we have made in the previous fits in order to stabilize the 
determination of
parameters we assume in addition that $a_{8}=0.51$ (with
$0.1$ as fictive theoretical error).

We get the following values of our parameters (describing the polarized 
quark distributions in eqs.(6-9)) from the fit to all existing data for spin 
asymmetries inclusive and semi-inclusive:
\begin{equation}
\begin{array}{lll}
a_{1}=\hspace*{0.293cm} 1.07,&a_{2}=-4.15,&a_{3}=\hspace*{0.101cm} 11.9,\\
b_{1}=-0.25,&b_{2}=\hspace*{0.36cm} 1.02,&b_{3}=\hspace*{0.350cm}3.18,\\
c_{u}=-0.90,&c_{d}=\hspace*{0.35cm} 3.44,&c_{s}=\hspace*{0.045cm}-3.90,\\
c=\hspace*{0.170cm}-0.37,&d=\hspace*{0.170cm}-0.04.&\\
\end{array}
\end{equation}
For this fit we get $\chi^{2}/N_{DF}=1.12$
We will not present the comparison of our fit with the
experimental inclusive asymmetries for proton, neutron and deuteron
targets because they are not very much different from the
previous figures. The comparison of the semi-inclusive spin
asymmetries for positive and negative hadrons on proton and
deuteron targets with experimental points obtained by SMC \cite{SMC} and
EMC \cite{emc} groups in CERN are given in figures 1a,b and 2a,b. 
Polarized quark
distributions for up and down valence quarks and up, down and
strange sea quarks are presented in figures 3a-3f.
The obtained quark distributions lead
to the following integrated quantities:
$\Delta u=0.74$ ($\Delta u_{v}=0.71$), $\Delta d=-0.52$
($\Delta d_{v}=-0.39$) and $\Delta s=-0.14$.
These numbers lead to the following predictions:
 $\Gamma_{1}^{p}=0.125$, $\Gamma_{1}^{n}=-0.084$, $a_{3}=1.25$, 
$\Delta\Sigma=0.08$, $\Delta M=-0.24$
The total sea contribution was strongly reduced in comparison
with the model considered before.
We have small positively polarized sea for up quark and stronger
negatively polarized sea for down and strange quarks.
As was already stressed in \cite{bt,bt1} the behaviour at small $x$
of our polarized quark distributions is determined by
the unpolarized ones and these do
not have the expected theoretically Regge type behaviour (which is also used 
by experimental
groups to extrapolate results to small values of $x$).  Unfortunately, 
some of the quantities change rapidly for $x<0.003$ (e.g. down quark
distribution). We present quantities
integrated over the region from $x=0.003$ to $x=1$ (it is practically 
integration over  the region which is covered by the experimental data, 
except of noncontroversial extrapolation for highest $x$). 
The corresponding quantities
are: $\Delta u_v=0.66$, $2\Delta\bar{u}=0.03$, ($\Delta u=0.69$), 
 $\Delta d_{v}=-0.29$, $2\Delta\bar{d}=-0.08$, ($\Delta d=-0.37$),
$\Delta s=-0.12$, $\Delta M=-0.17$ and $\Gamma_{1}^{p}=0.123$, 
$\Gamma_{1}^{n}=-0.054$, $a_{3}=1.06$, $\Delta\Sigma=0.20$.   

The above given values extrapolated to $x=0$ using Regge
type behaviour $x^{-0.25}$ for small $x$ give:
 $\Delta u_v=0.70$, $2\Delta\bar{u}=0.02$, ($\Delta u=0.72$), $\Delta
d_{v}=-0.33$, $2\Delta\bar{d}=-0.10$, ($\Delta d=-0.43$), $\Delta
s=-0.13$, $\Delta M=-0.21$ and $\Gamma_{1}^{p}=0.126$, 
$\Gamma_{1}^{n}=-0.066$, $a_{3}=1.07$, $\Delta\Sigma=0.16$,

Our results can be compared with that for the non strange sea
polarization $\Delta \bar{q}=-0.02$ obtained by SMC group
\cite{SMC} (under the assumption  $\Delta \bar{u}=\Delta \bar{d}$
and assumed value for $\Delta s=-0.12$). Also our valence up and
down quark contributions are different from those presented in \cite{SMC}.

When we use this model (with flavour dependent
polarization of the sea contributions) to make a fit to the
subset of the data on inclusive spin asymmetries we get the
fitted parameters and results very close to the considered model
with $\chi^{2}/N_{DF}=1.01$.    
For example we get $2\Delta\bar{u}=0.05$, $2\Delta\bar{d}=-0.09$ and $2\Delta
\bar{s}=-0.12$ in $0.003\leq x \leq1 $ region. So it was
possible to obtain the information about flavour dependence of 
polarized sea without taking into account data on semi-inclusive hadron spin
asymmetries only. That shows a 
consistency of both sets of data (inclusive and semi-inclusive)
with our assumption of flavour dependence of polarized sea.
When we use instead of the MRS fit used in this paper 
another version called A' \cite{mrs2} the similar results are
obtained. The results practically do not depend on the precise
form of the MRS fit for unpolarized parton distributions used as
a starting point for polarized parton distributions.
As in the previous papers \cite{bt,bt1} we have also tried to
include polarized gluons along the line of
\cite{ros} assuming for the gluon distribution:
\begin{equation}
\Delta G(x)=x^{-0.801}(1-x)^{6.06}(d_{1}+d_{2}\sqrt{x}),
\end{equation}

\noindent with a new $d_{1}$ and $d_{2}$ constants which have to
be fitted.
The appearance of non-zero gluonic distribution affects our formulas 
only through the substitution: 
$\Delta q \Rightarrow \Delta q-\frac{\alpha_{s}}{2\pi}\Delta G$.
In such a fit we got (after integration)
the negative sign of the gluonic 
contribution, i.e. opposite to the one expected from the paper
\cite{ros}. Our conclusion is that the inclusion of gluon
contribution does not lead to the substantial improvement of the
fit. 
Encouraged by the present results we can make the next step and
look for the differences in the structure of polarized sea for
quarks and antiquarks (semi-inclusive hadron spin asymmetries
differentiate these contributions). Similar suggestions were
coming from model considerations \cite{brod1}.
As before we assume:
\begin{equation}
\Delta M_{i}(x)=x^{-0.567}(1-x)^{10.1}(c+c_{i}\sqrt{x}),
\end{equation}
where i labels this time up, down, strange and charm quarks and up and
down antiquarks. We have:
\begin{eqnarray}
\Delta u_{M}(x)&=&0.196\Delta M_{u}(x)-0.5\Delta \delta (x), \nonumber \\
\Delta\bar{u} (x)&=&0.196\Delta M_{\bar{u}}(x)-0.5\Delta
\delta(x), \nonumber \\
\Delta d_{M}(x)&=&0.196\Delta M_{d}(x)+0.5\Delta \delta (x), \nonumber \\
\Delta\bar{d} (x)&=&0.196\Delta M_{\bar{d}}(x)+0.5\Delta
\delta(x),  \nonumber \\
2\Delta\bar{s} (x)&=&0.196\Delta M_{s}(x), \\
2\Delta\bar{c} (x)&=&0.020\Delta M_{s}(x). \nonumber
\end{eqnarray}

The data on semi-inclusive hadron spin asymmetries can not
distinguish between $q$ and $\bar{q}$ for strange and charm
quarks. The obtained fit is very similar to our basic fit. With two
new parameters we have got $\chi^2$ only 0.5 smaller than
before. The integrated quantities are very close to that gotten from
eq.(8). The obtained sea contributions integrated in $0.003\leq
x \leq 1$ region are:
$\Delta u_{M}=-0.04$, $\Delta\bar{u}=0.09$, $\Delta
d_{M}=-0.02$, $\Delta\bar{d}=-0.07$ and $\Delta s=-0.12$.

It seems that with the data on semi-inclusive hadron asymmetries
it is difficult to make definite
statement about quark antiquark structure of polarized sea
because all non-strange sea contributions are rather small and
their signs depend on whether we use to fit all data points
namely SMC and EMC or only SMC (of course together with
inclusive data). More precise data are needed to make conclusive
statement. When we use MRS fit A' the results practically do not
change. The summed contributions for quark and antiquark are very
close to the results obtained before and do not strongly depend
whether we take for fitting SMC data only or SMC and EMC data.

We have extended our determination of polarized parton
distributions by taking into account not only all available
inclusive spin asymmetries but also semi-inclusive hadron spin
asymmetries on protons and deuterons measured by experimental
groups in CERN. By relaxing our previous assumption about the
polarized sea we have found polarized sea contributions for up,
down and strange quarks. The total sea contribution was strongly
reduced. The results show stability and do not
change when we make fit to inclusive data only or to all data
(inclusive and semi-inclusive). There is also a very small change
in calculated parameters when we use the other version of the
MRS fit (called A') \cite{mrs}. The attempt to determine
quark-antiquark differences in the polarized sea structure do
not seem to be conclusive because we have to do with the small numbers
and the signs of different contributions depend on the subset of
data used for fitting. The gluon
contribution have sign opposite to the expected one and
does not seem to influence the fit significantly.

\newpage 
{\bf Figure captions}

\begin{itemize}
\item[ Figure 1 \ ] The comparison of semi-inclusive spin
asymmetries for production of positive (a) and negative (b) hadrons
on protons with the curve gotten from our fit.
Points are taken from SMC \cite{SMC} and EMC \cite{emc} experiments. 
\item[Figure 2 \ ] The comparison of semi inclusive spin asymmetries 
for production of positive (a) and negative (b) hadrons
(SMC data) with the curve gotten from our fit.
\item[Figure 3 \ ] Our predictions for polarized quark
distributions for : valence a) u, b) d quarks and sea c) u, d) d, e) s, 
f) (u+d)/2 quarks.
\end{itemize}
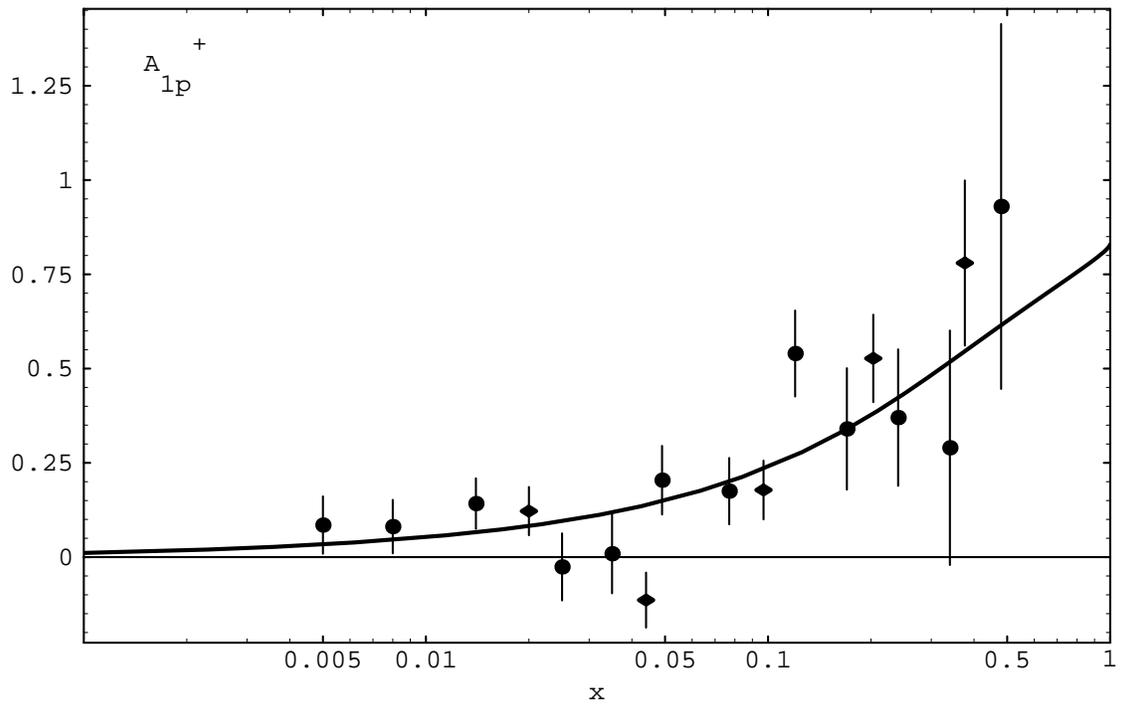
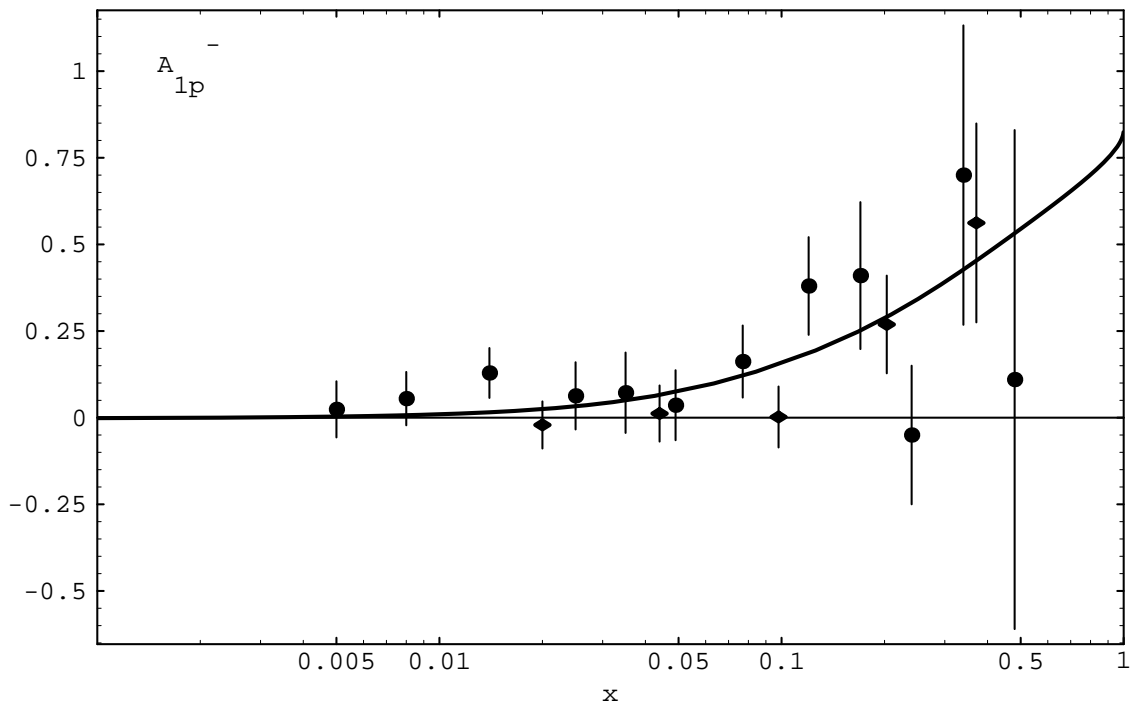
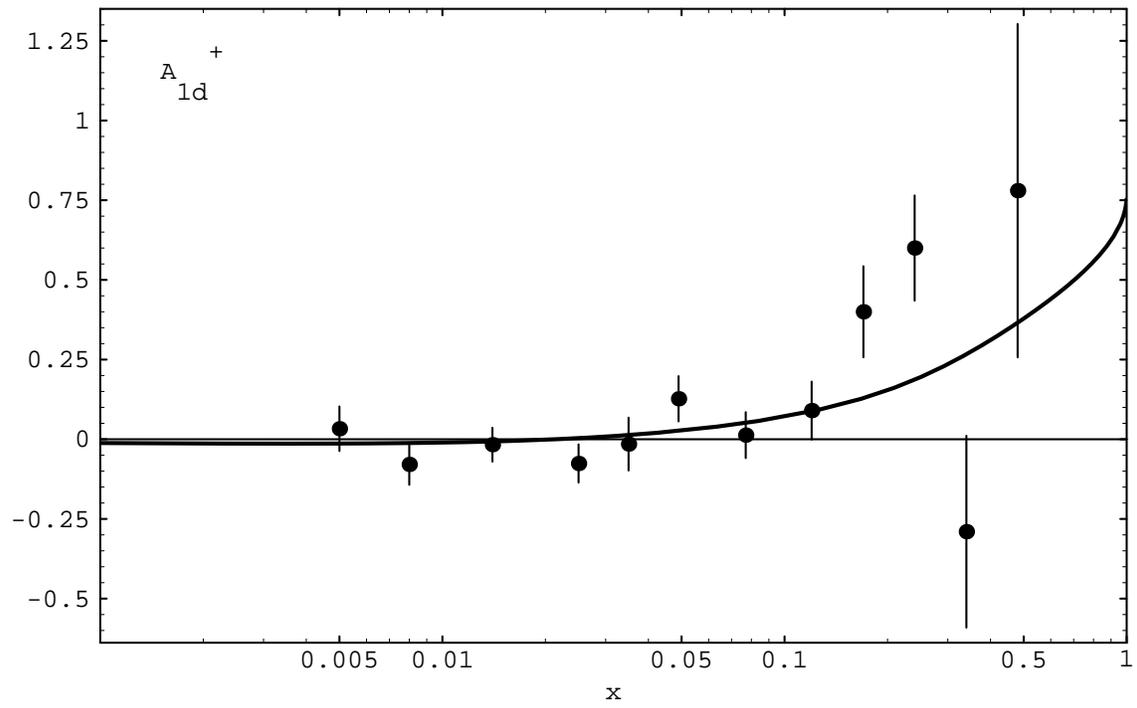
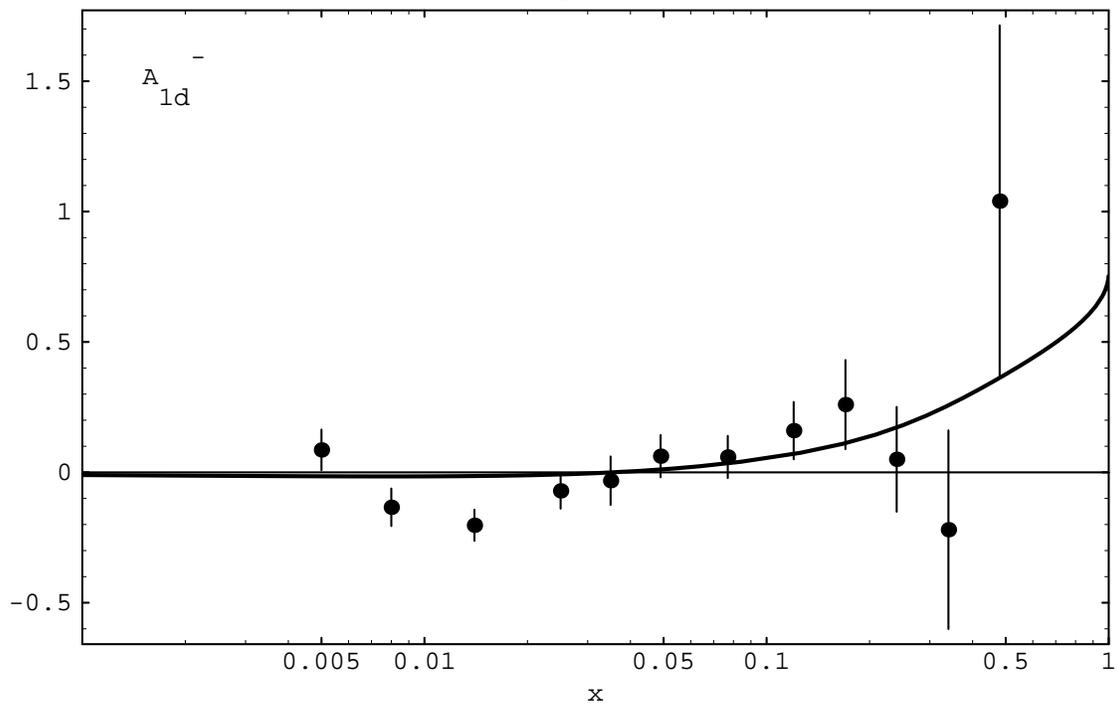
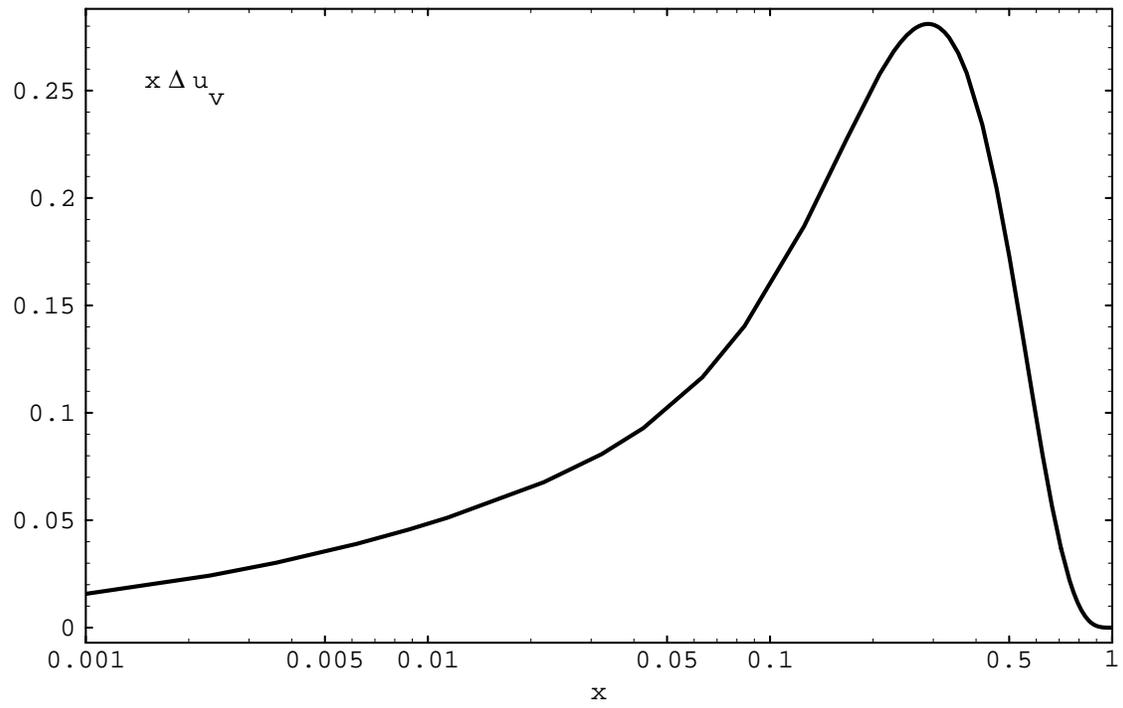
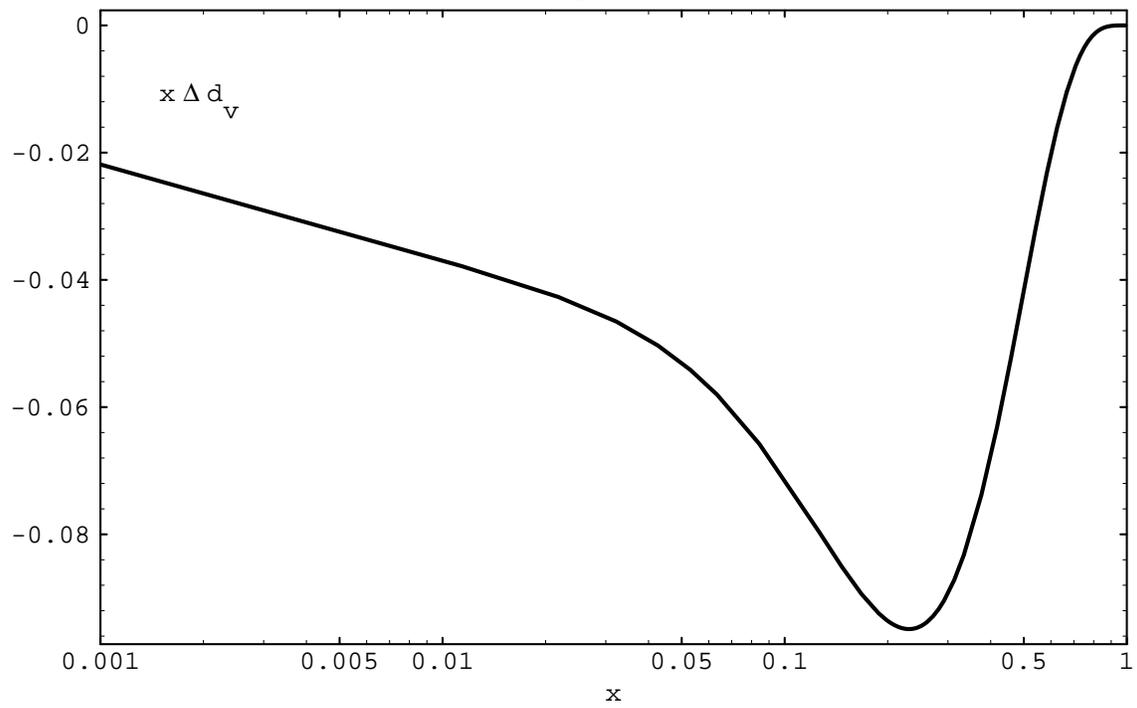
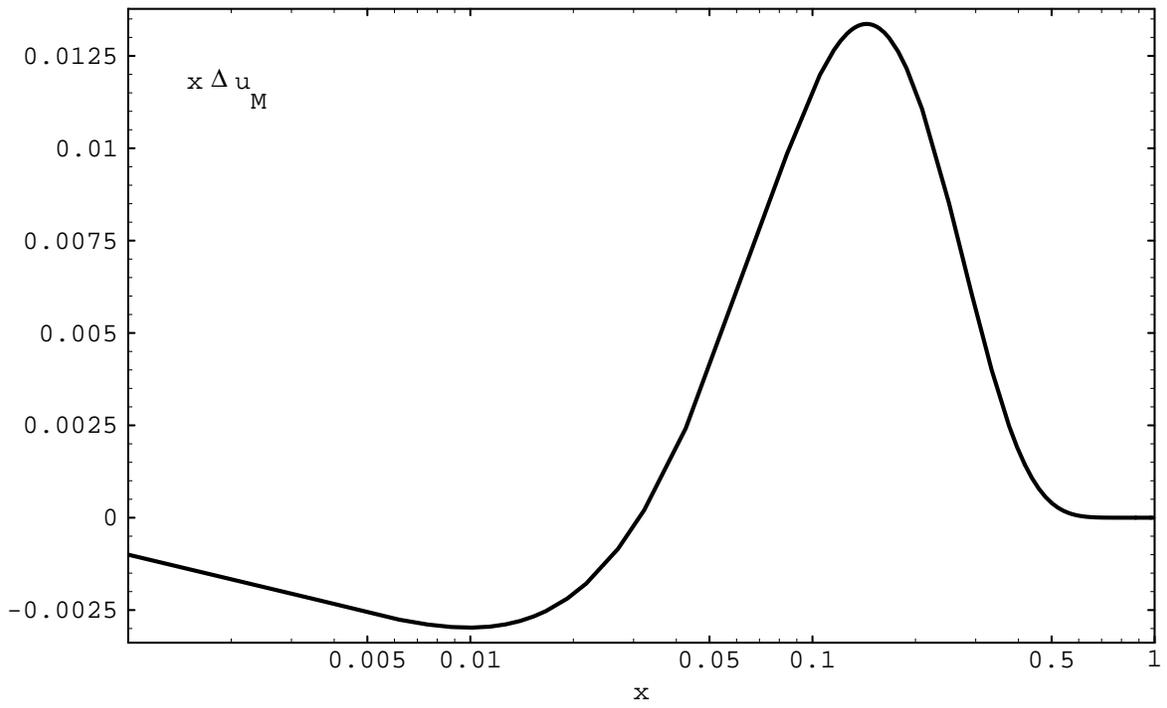
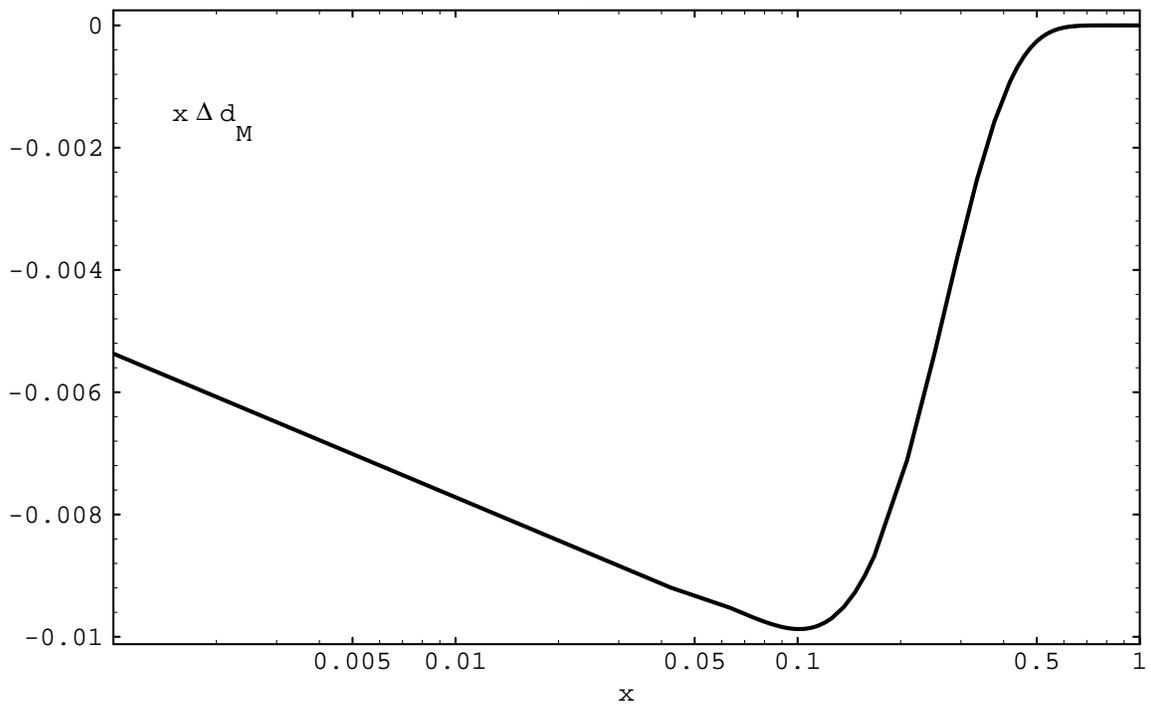
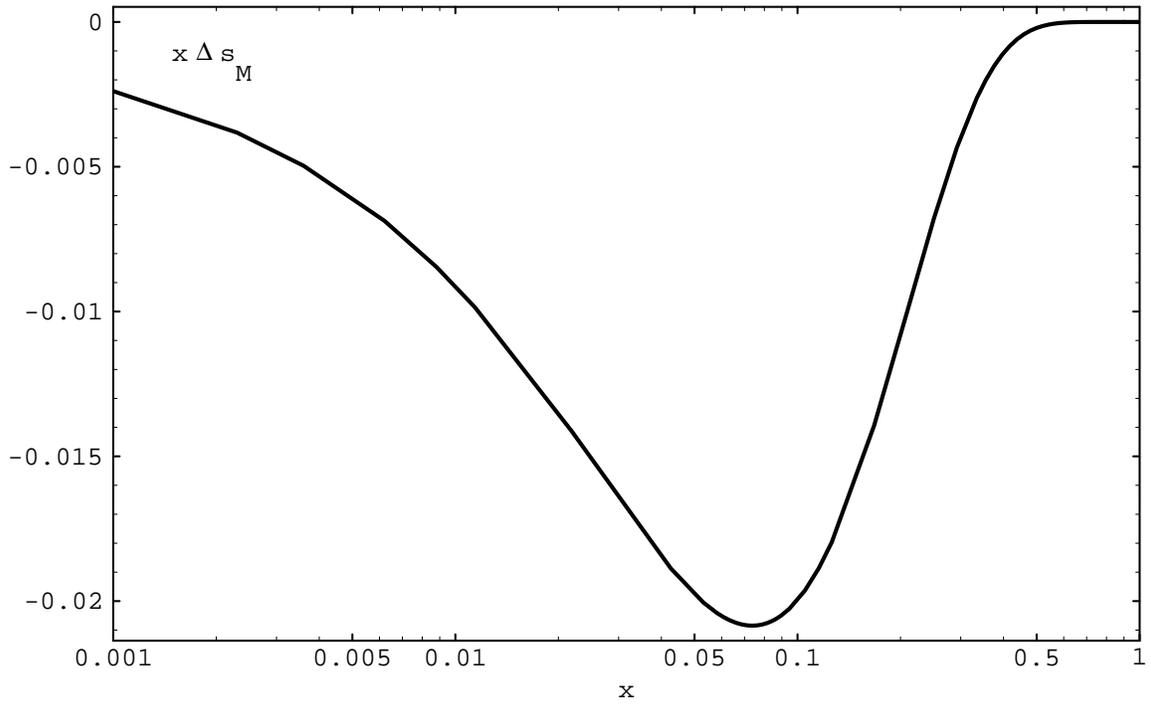
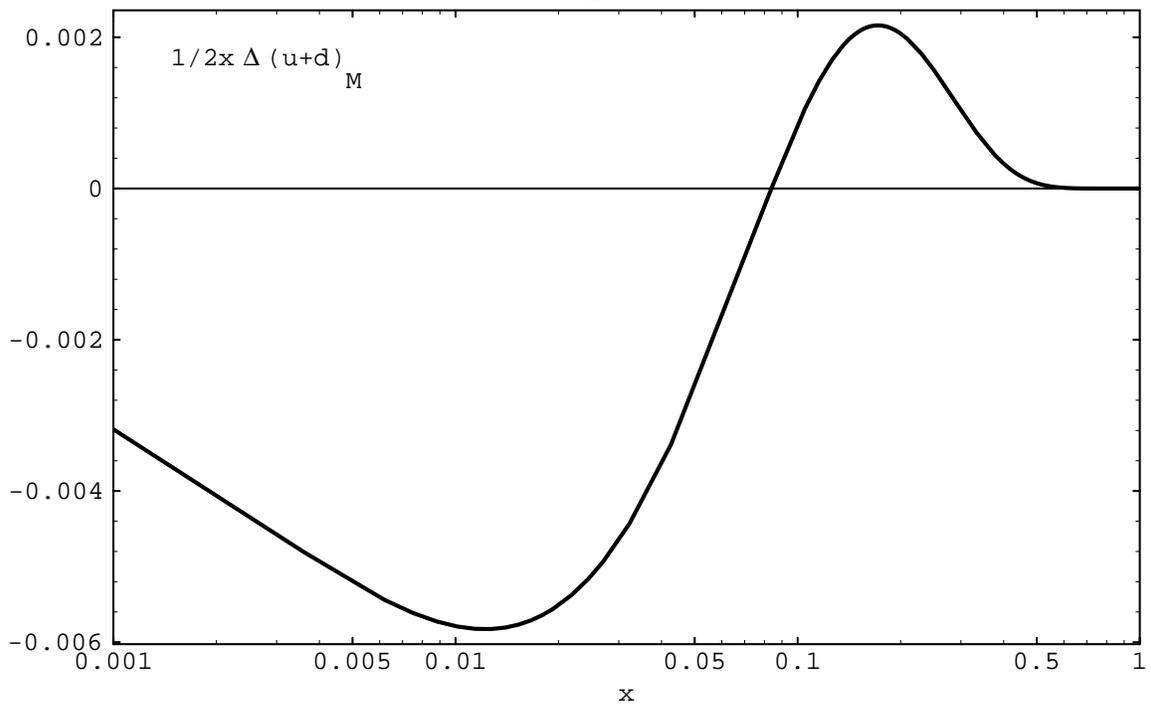
\end{document}